\documentclass[10pt,conference]{IEEEtran}
\IEEEoverridecommandlockouts

\usepackage{cite}
\usepackage{amsmath,amssymb,amsfonts}
\usepackage{algorithmic}
\usepackage{graphicx}
\usepackage{textcomp}
\usepackage{xcolor}
\usepackage{microtype}
\usepackage{array}
\usepackage{cellspace} 
\usepackage{hyperref}
\usepackage{blindtext}

\def\BibTeX{{\rm B\kern-.05em{\sc i\kern-.025em b}\kern-.08em
    T\kern-.1667em\lower.7ex\hbox{E}\kern-.125emX}}
\begin{document}

\title{Extending Behavioral Software Engineering: Decision-Making and Collaboration in Human-AI Teams for Responsible Software Engineering\\
{\footnotesize}
\thanks{\rule{8.5cm}{0.4pt}\\Submission to the Doctoral and Early Career Symposium of the Cooperative and Human Aspects of Software Engineering 2025 conference. © with the authors.}
}

\author{
    \IEEEauthorblockN{Lekshmi Murali Rani}
    \IEEEauthorblockN{
        \textit{Chalmers University of Technology and University of Gothenburg}\\
        SE - 41296, Gothenburg, Sweden \\
        Email: lekshmi@chalmers.se
    }
}

\maketitle

\begin{abstract}
The study of behavioral and social dimensions of software engineering (SE) tasks characterizes behavioral software engineering (BSE); however, the increasing significance of human-AI collaboration (HAIC) brings new directions in BSE by presenting new challenges and opportunities. This PhD research focuses on decision-making (DM) for SE tasks and collaboration within human-AI teams, aiming to promote responsible software engineering through a cognitive partnership between humans and AI. The goal of the research is to identify the challenges and nuances in HAIC from a cognitive perspective, design and optimize collaboration/partnership (human-AI team) that enhance collective intelligence and promote better, responsible DM in SE through human-centered approaches. The research addresses HAIC and its impact on individual, team, and organizational level aspects of BSE.
\end{abstract}


\section{Introduction}
Software Engineering (SE) is a decision-intensive process that includes multiple attributes, criteria, and objectives. The decisions in SE involve problem-solving in a dynamic environment where the decisions are based on very limited available information, unclear and contradictory goals, time and resource constraints~\cite{Burge2008}. In general, decision-making (DM) involves selecting the best alternative (from several alternatives) that is sufficiently differentiated from its closest competitors~\cite{takemura2021behavioral,svenson1992differentiation}. 
With the integration of AI in the workflow, the DM capabilities of humans can be enhanced where the AI can support the humans in decision-making by analyzing large amounts of data to identify hidden patterns~\cite{heilig2023decision}. While this human-AI collaboration (HAIC) in DM can make remarkable changes in SE decision-making, the need for informed, smart, and value-based decisions~\cite{mendes2018towards} can benefit from responsible DM to ensure responsible software engineering. The successful and responsible implementation of any software can be traced back to the decisions that were made at an earlier point in time in the software engineering life cycle~\cite{cunha2016decision}, with effective DM benefiting from adaptive and symbiotic HAIC~\cite{xu2020artificial,fragiadakis2024evaluating}. 

Despite the potential of AI in complex problem-solving and DM skills, integrating AI as a collaborative team member with humans brings cognitive, behavioral, and organizational challenges. This research investigates these challenges to optimize HAIC for responsible and effective DM in SE. 

\section{The problem}
Behavioral Software Engineering (BSE) explores the behavioral and social aspects of SE at individual, team, and organizational level~\cite{lenberg2014towards}. While traditional BSE has primarily focused on human teams, the integration of AI as a collaborative partner introduces the human-AI team-level concept in the context of BSE and its subsequent influences on the three units of BSE~\cite{collins2024building}.  The human-AI team concept extends existing BSE sub-units of analysis by addressing:

\begin{itemize}
    \item How AI influences cognitive, behavioral, and socio-emotional aspects of SE professionals during collaboration.
    \item How to optimize human-AI collaboration for effectively performing SE tasks and DM.
    \item How organizations can integrate and govern HAIC to promote responsible and ethical SE practices.
\end{itemize}

Existing studies give less emphasis on the dynamic nature of the human-AI team, their collaboration with an emphasis on human aspects, and their impact on the complex and highly context-dependent field of SE. This gap in research and practice has resulted in the limited adoption of AI as a collaborative and symbiotic partner in SE tasks~\cite{russo2024navigating,agrawal2024artificial}. This research aims to address these gaps by investigating how to optimize HAIC for responsible and effective DM in SE.

\section{Research Objectives}
The research aims to identify the challenges and key factors that influence human-AI collaboration, designing and optimizing HAIC, and then devising organizational strategies to integrate and regulate HAIC.

The first phase of the research explores the cognitive, behavioral, and socio-emotional aspects of HAIC. The key areas of exploration include:
\begin{itemize}

    \item Explore how professionals (eg, project managers, software engineers, requirements engineers) collaborate with AI for DM and how AI influences their cognitive load, trust, and biases.
    \item Identify the psychological and socio-emotional factors that impact individual DM during HAIC and devise mitigation strategies for the socio-emotional gap in the human-AI team.
    \item Analyze how software professionals adapt to AI as a cognitive partner in collaboration.
    \item Explore the unique challenges faced by software professionals during HAIC that are not observed in human-human collaboration.
\end{itemize}

The next phase of the research plans to focus on designing and optimizing HAIC by focusing on a human-centered approach. This stage follows a human-centered approach to achieve human-AI synergy in DM. The key areas of exploration include:
\begin{itemize}

    \item Explore the key factors that influence the effectiveness of the HAIC.
    \item Design a collaboration model for a human-AI team as a framework to optimize and improve the SE tasks and DM process in collaboration. 
    \item Develop a guideline for interaction patterns that suggests when AI should lead and when humans should lead in a collaborative DM setup.
    \item Develop an effective strategy for creating a mental model during HAIC for enhanced decision-making.
\end{itemize}

The final phase of the research plan is to address the organizational aspect of human-AI collaboration and devise governance strategies to make human-AI collaboration more effective and part of organizational culture. Stages include:
\begin{itemize}
    \item Study the impact of HAIC on organizational DM culture.
    \item Develop organizational strategies and best practices to promote effective collaboration and responsible DM through HAIC.
\end{itemize}

\section{Research Questions}
The research questions for this PhD study are framed around the research objectives and BSE levels.

\textbf{RQ1:} \textit{What are the key challenges and influential factors affecting HAIC in SE, particularly in SE tasks and decision-making processes?}
     \begin{itemize}
         \item \textbf{RQ1.1:} How does HAIC influence individual DM styles when compared to human-human collaboration?
         \item \textbf{RQ1.2:} What cognitive, behavioral, socio-emotional, and ethical challenges arise from HAIC?
     \end{itemize}

\textbf{RQ2:} \textit{How can HAIC be designed and optimized to ensure effective collaboration and responsible DM in SE?}
    \begin{itemize}
        \item \textbf{RQ2.1:} What key factors influence HAIC effectiveness in SE decision-making?
        \item \textbf{RQ2.2:} How can HAIC be designed to enhance collective intelligence and make responsible, unbiased, and value-based decisions?
    \end{itemize}

\textbf{RQ3:} \textit{What strategies and governance frameworks are required to successfully create HAIC at the organizational level while ensuring responsible and ethical DM?}
    \begin{itemize}
        \item \textbf{RQ3.1:} How does HAIC influence and transform organizational culture?
        \item \textbf{RQ3.2:} What best practices or governance strategies can support responsible HAIC at the organizational level?
    \end{itemize}

\section{Methodology}
Being in the first year of PhD, the current research uses knowledge-seeking research that aims to increase the understanding of HAIC, with plans for applied research in the subsequent phases. The studies plan to use inductive logic to find connections in the observed data and thus build theory and generalizations from them. The research design will be constructivist/interpretivist and utilize both qualitative and quantitative approaches to address the research questions~\cite{hoda2024}. As the research field is BSE, qualitative approaches would help to capture the facts as perceived by the individuals (subjective) based on the specific context in depth while quantitative methods will bring more generalizability and rigor in the study~\cite{wohlin2015towards}. RQ1 will mainly use literature reviews, surveys, and interviews. RQ2 will mainly use experiments and design science approaches. RQ3 mainly uses case studies, experiments, and design science approaches. The research design canvas based on the current research plan can be found in \href{https://doi.org/10.5281/zenodo.14743120}{Zenodo}~\cite{rani2025advancing}.

\section{Contribution and Potential Impact}
\begin{figure}[ht]
    \centering
    \includegraphics[width=1\linewidth]{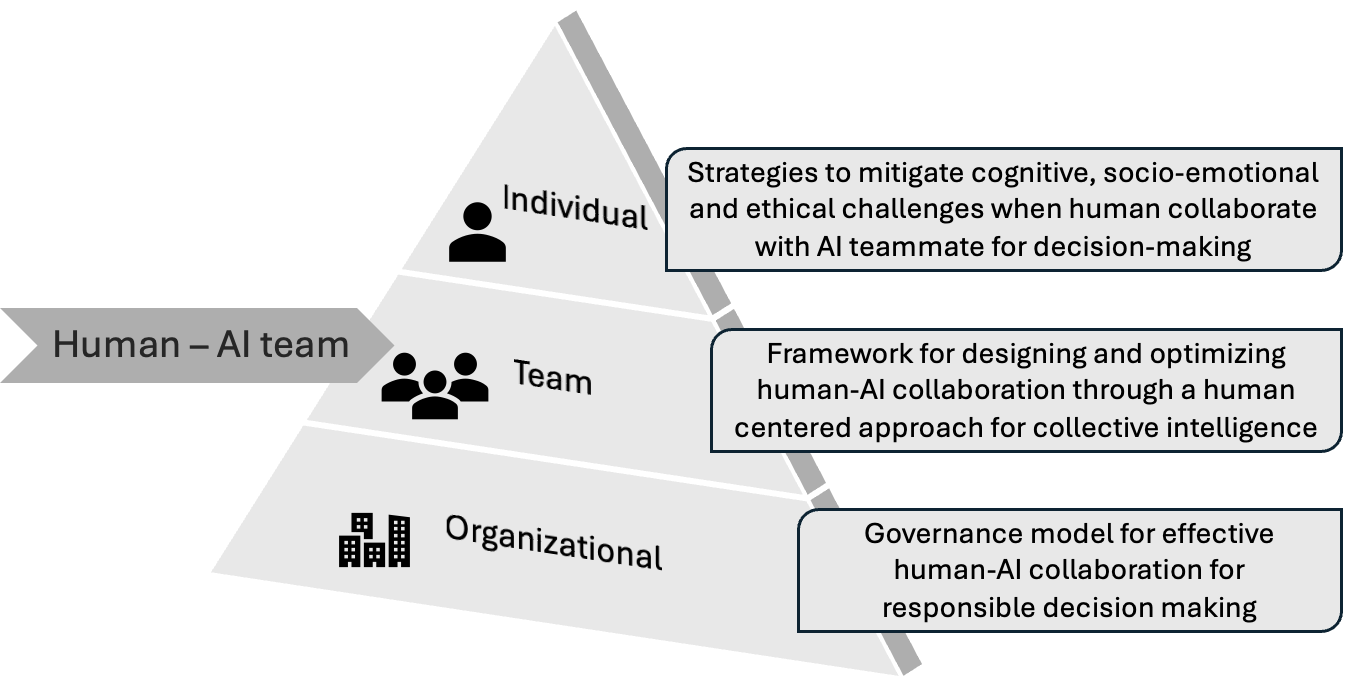}
    \caption{Research Contribution}
    \label{contribution}
\end{figure}

In this PhD research, the main contribution will be the introduction of the human-AI team concept in BSE, where the AI is considered as a cognitive partner with roles and responsibilities. By adding the perspective of the human-AI team to the BSE level, the research explores how HAIC influences the individual, team, and organizational level aspects of BSE. This research mainly sets the stage for the evolution of human-AI partnerships, focusing on more behavioral aspects that could enhance software engineering DM. The study contributes to the software engineering field by answering the key questions(simplified in a popular science approach) in Figure \ref{contribution}.

%\bibliographystyle{IEEEtran}
%\bibliography{mybibfile} 

\end{document}